\documentclass{elsart}

\usepackage{amssymb}
\journal{Progress in Surface Science}
\usepackage{graphicx}
\begin{document}

\begin{frontmatter}

\title{Quantum size effects in the low temperature layer-by-layer 
growth of Pb on Ge(001)}

\author[tasc]{L. Floreano\corauthref{floreano}}
\ead{floreano@tasc.infm.it}
\corauth[floreano]{Tel: +39 040 3756442, +39 040 3758369; fax: +39 040 226767}
\author[tasc,unilju]{D. Cvetko}
\author[tasc,units]{F. Bruno}
\author[tasc,unilju]{G. Bavdek}
\author[tasc]{A. Cossaro}
\author[tasc]{R. Gotter}
\author[tasc]{A. Verdini}
\author[tasc,units]{A. Morgante}

\address[tasc]{Laboratorio TASC-INFM, Basovizza SS-14 Km. 163.5, 
I-34012, Trieste, Italy.}
\address[unilju]{also at: J.Stefan Institute, Physics department, University of Ljubljana, Slovenia.}
\address[units]{also at: Physics department, University of Trieste, Italy.}

\begin{abstract}

The electronic properties of thin metallic films deviate from the 
corresponding bulk ones when the film thickness is comparable with 
the wavelength of the electrons at the Fermi level. This phenomenon, 
referred to as quantum size effect (QSE), is also expected to affect 
the film morphology and structure leading to 
the ``electronic growth'' of metals on 
semiconductors. 
Shuch effect may be observed when metals are grownon substrates held at low 
temperature and are manifested through the occurrence of  ``magical'' 
thickness islands or critical thickness for layer-by-layer growth.
In particular, layer-by-layer growth of Pb(111) films has been 
reported for deposition on Ge(001) below $\sim$~130~K.
An extremely flat morphology is preserved throughout deposition 
from four up to a dozen of monolayers. 
These flat films are shown to be metastable 
and to reorganize into large  clusters uncovering the first Pb layer 
pseudomorphic to the underlying Ge(001) substrate already at room 
temperature.
Indications of QSE induced structural variations of the 
growing films have been  reported for Pb growth on both Si(111) and 
Ge(001). In the latter case,  
the apparent height of the Pb(111) monatomic step was shown to change 
in an oscillatory fashion by He atom scattering (HAS) during 
layer-by-layer growth at low temperature.
The extent of the structural QSE has been obtained by a comparison 
of the HAS data with X-ray diffraction (XRD) and reflectivity  
experiments. Whereas step height variations as large as 20\%~ have been
measured by HAS reflectivity, the displacement of the atomic planes 
from their bulk position, as measured by XRD, has been found to mainly 
affect the topmost Pb layer, but with a lower extent, i.e. the QSE
 observed by HAS are mainly due to a 
perpendicular displacement of the topmost layer charge density. 
The effect of the variable surface relaxation on the surface 
vibration has been studied from the acoustic 
dispersion of the low energy phonons, as measured by inelastic HAS.
\end{abstract}


\end{frontmatter}


In the field of surface science, a strong experimental and theoretical 
effort has been traditionally devoted to the study of the growth 
and diffusion processes for their relevance in the fabrication of 
novel electronic devices. From an experimental point of view, the main 
issue has been to achieve the conditions for layer-by-layer growth 
or a regular patterning of the growing film in order to obtain sharp 
interfaces between different materials.
In particular much attention has been drawn to a few key 
cases of the homoepitaxial growth (such as reentrant layer-by-layer growth,  
inverse growth, ion assisted growth) 
as well as heteroepitaxial growth (particularly surfactant assisted growth) 
giving rise to a consistent thermodynamical frame for describing the growth 
and diffusion processes \cite{NATO}.  

While in the past most of the attention was dedicated to high temperature 
film growth of semiconductors (to fabricate electronic devices), 
at present the deposition of metals at low substrate temperature is the 
most attractive field for investigating the correlation between 
the structural and electronic properties of the film. 
In heteroepitaxial films, the suppression of thermally 
activated processes, such as surface diffusion and intermixing, 
enables the formation of two-dimensional (2D) 
metal films with sharp interfaces. Thus, 
the interplay between the structural and electronic lengthscales 
of the metallic film can be put in evidence.
In addition, low temperature 
growth of metal on metal surfaces is also increasingly employed in the 
study of magnetic systems thanks to the possibility of inhibiting intermixing 
processes.

Typically, electronic effects are not observed upon  
deposition on semiconductors or insulators at room 
temperature (RT). 
In fact, most metals display a cluster growth behavior, 
either in a Wolmer-Weber, or in a Stranski-Krastanov mode. 
The thermal energy at the metal/semiconductor interface 
is usually high enough to allow metal adatoms to migrate uphill
at the island step edges, thus to explore various thermodynamical 
configurations. Both the tendency to reduce the 
interface area and that to reduce the ratio between the cluster area 
and its volume may favor the formation of three-dimensional (3D) 
metal clusters. Any electronic 
effect due to size confinement is then smeared out by the 
inhomogeneous spatial and 
size distribution of the clusters. 
A possible path for growing films with a smoother morphology is to 
inhibit the uphill flow of the deposited atoms 
by lowering the substrate temperature.
This route has been shown 
to allow a certain degree of control in the morphology of the growing 
films leading to the formation of regular patterns of islands with 
homogeneous size and shape, or even to layer-by-layer growth.
Such controlled non-equilibrium morphology can display the effects originated by 
the quantum confinement of the metal electrons.
These phenomena, 
referred to as quantum size effect (QSE), are clearly manifested by the 
appearance of quantum well states (QWS) in thin metal films and 
nanoislands \cite{handbook,chiang,milun}, as well as by the spin density waves 
observed in heterogeneous magnetic systems \cite{johnson}, which give 
rise to the exchange magnetic coupling 
in layered systems \cite{ortega,stiles,bruno}. 
QSE were first observed by Jaklevic and coworkers, who measured the 
tunnelling current of thin Pb electrodes 
\cite{jaklevic}, while the QWS were directly observed 
by valence band photoemission from single layer metal films \cite{wallden}. 
Later on,  QWS have been measured by photoemission
in several metal films on both metal \cite{kawakami,paggel} 
and semiconductor substrates \cite{devans,arranz,aballe}. 

In general, when the 
thickness $D$ of a metal film 
is comparable with the Fermi wavelength $\lambda _{F}$ 
of the film electron gas, cyclic variations of the surface charge density and 
work function are expected as the film thickness 
approaches an odd multiple of $\lambda _{F}/4$ \cite{schulte}.
As new layers are 
added to the metal film, cyclic variations of the perpendicular displacement 
of the surface charge density should be observed.
Among different metals, Pb is certainly a suitable candidate to 
display strong QSEs, since its bulk interlayer spacing $d_{Pb} = 2.86$~\AA, 
along its preferred growth orientation [111], is $\sim \frac{3}{4} 
\lambda_{F}$ \cite{lambdaPb}. Deviations of the physical properties in 
thin Pb films from the bulk ones are thus expected with a bi-layer 
periodicity.

The first evidence of such oscillatory behavior dates 
back to the works by Jalochowski and Bauer, who measured the 
electrical resistivity \cite{jalochowski,jalochowski1}, 
valence band photoemission \cite{jalochowski2} 
and the RHEED reflectivity \cite{jalochowski,jalochowski3} during Pb deposition 
on Si(111)-(7$\times$7) and Si(111)-Au(6$\times$6) at low temperature. 
Double layer modulations, characteristic of QSE, were observed in both 
resistivity and RHEED, with their amplitude increasing with the quality 
of the growing film surface (domain size and flatness). Unexpectedly, 
the best growth conditions were achieved at the lowest temperatures 
down to 16~K. In this temperature range, no kinetic mechanisms are 
activated and other mechanisms must be involved in the diffusion 
process. In fact, it was also observed that variations of 
the growth rate by a factor of 20 were yielding the same shape of the RHEED 
oscillations, i.e. the same film morphology \cite{jalochowski3}.
The same system was later studied by X-ray diffraction (XRD) and 
compared to He atom scattering (HAS) and RHEED 
reflectivity \cite{schmicker,edwards}. Interestingly, Hibma and 
coworkers found that the  crystalline structure of Pb(111) is only 
observed after the deposition of 5 ML of Pb \cite{edwards}, 
the same thickness also 
corresponds to the onset of an almost layer-by-layer 
growth \cite{schmicker}. 

The existence of a critical thickness for the formation of a flat 
surface appears to be characteristic of the low temperature growth of 
metal on semiconductors, having been also observed for Pb on Ge(001) 
with HAS \cite{crottini} and for Ag on Si(111) \cite{huang}, 
GaAs(110) \cite{smith,yu}, and other III-V compounds \cite{chao} with 
STM. Below the critical thickness or at intermediate 
substrate temperature, the metal 
clusters have been found to display a regular prismoidal shape with a 
flat surface. These flat islands display a preferred 
thickness, i.e. only bi-layer islands are observed for Ag on Si(111) 
\cite{gavioli}, while odd $n$-layer islands ($n = 5, 7, 9$) are 
favored for Pb on Si(111) \cite{budde,hupalo1,hupalo2}.
In the latter case, Tringides and coworkers demonstrated the 
possibility to grow Pb islands with homogeneous thickness by properly 
selecting the substrate temperature and Pb coverage \cite{hupalo2}, 
or even by changing the substrate reconstruction upon 
pre-deposition \cite{hupalo1,yeh}.

The stability of these flat morphologies has been recently 
discussed on the basis of a detailed balance among the energetics 
of the charge quantum confinement, the charge transfer at the interface, 
the Friedel's charge density oscillation  
and the stress at the interface \cite{zhang,suo}. 
These competing phenomena can lead to the manifestation of
the ``electronic growth'' of metals on semiconductors, i.e. to the 
existence of a critical thickness for the onset of 
layer-by-layer growth and/or magic islands. 

These flat ``magic'' islands 
represent an almost ideal quantum box for the metal 
Fermi electrons and QSE can be observed by means of a local probe such 
as STM. Altfeder measured by STM the electron interference fringes on 
the flat surface of a Pb
quantum wedge grown on a staircase of Si(111) terraces \cite{altfeder}.
In a similar fashion, an oscillatory variation of the 
monatomic step height as a function of the film thickness 
was observed in Pb layer-by-layer growth on a flat Ge(001) 
surface by He atom reflectivity, i.e. by a 
probe sensitive only to the surface charge density \cite{crottini}.
The apparent step height was found to oscillate around the value 
of the Pb(111) bulk interlayer spacing with an amplitude up to 20\%. 
According to theoretical calculations, these variations should be 
accompanied by the relaxation of the topmost film layer, so that 
the structural variations of 
the film partly counteract the extent of the electronic 
property variations \cite{feibelman}.
More detailed calculations indicate that 
all the interlayer spacings of the growing film 
should be affected by QSE \cite{batra,cho}, 
in such a way to further minimize the total energy.
In fact, indirect evidence of a structural effect has been recently 
found by Tsong via STM spectroscopy on Pb magic islands, where 
the deviation of the apparent step height, 
has been studied as a function of the tip bias voltage. 
Furthermore, the energy separation between the highest filled states 
and the lowest empty ones  
has been studied as a function of the island thickness  
to assign the quantum number to the QWS within the islands \cite{su}.

In the following section, the low 
temperature deposition of Pb on Ge(001) is characterized
by both HAS and X-ray reflectivity (XRR) measurements. 
The existence of a critical thickness for the formation of crystalline 
Pb(111) islands as well as for the onset of 
layer-by-layer growth is discussed.
An extremely flat morphology is observed throughout deposition 
from five up to a dozen of monolayers, then the number of exposed 
layers starts to increase. These flat films are shown to be metastable 
and to reorganize into large  clusters uncovering the first Pb layer 
pseudomorphic to the underlying Ge(001) substrate upon annealing to room 
temperature.

A direct measurement by X-ray diffraction of the topmost layer relaxation 
and inner layer spacing distribution is presented 
in the thickness range of layer-by-layer growth.
These results are compared with previous 
values of the monatomic step height measured by HAS.
Whereas step height variations up to 20\%~ have been
measured by HAS reflectivity, the displacement of the atomic planes 
from their bulk position, as measured by X-ray diffraction (XRD), 
is found to mainly 
affect the topmost Pb layer, but with a much lower extent, 
i.e. the QSE observed by HAS are mainly due to a 
perpendicular displacement of the topmost layer charge density. 
Finally, we will present the effect of the oscillating 
surface relaxation on the surface phonons of the Pb(111) film surface, 
as measured by the inelastic HAS.

\section{Experimental}

The experiments have been performed with the HAS apparatus of the 
I.N.F.M.-TASC National Laboratory \cite{apparato} 
and at the Elettra Synchrotron facility in Trieste 
(Italy) with the I.N.F.M. ALOISA beamline \cite{aloweb}.
Both experimental stations make use of the same type of manipulator (a 
6-degree of freedom manipulator with 0.01$^{\circ}$ precision on each 
rotation axis), sample holder, MBE cryo-panel and sputter gun, 
thus allowing us to easily exchange the samples between the two systems 
and to adopt the same preparation procedures. In the ALOISA end station,
X-ray photoemission is used to check for the substrate cleanness and 
a RHEED system is used to check for the surface long range order 
prior to Pb deposition. 
The experiments have been performed on a set of about a dozen of 
different Ge(001) samples obtained by different wafers and rods.
The substrate is cleaned by Ar$^{+}$ ion bombardment at 1~keV, while 
annealing the sample up to about 1000~K ($T_{m}$ = 1211~K) and  
monitoring the half-integer peaks of the $(2 \times 1)$ reconstruction. 
Both heating and sputtering are switched off 
at the occurrence of the $(2 \times 1) \rightarrow (1 \times 1)$ 
phase transition. Upon cooling below $\sim$~240~K, 
this procedure yields a well ordered $c(4 \times 2)$ structure
 with average domain size of $\sim$~1000~\AA~ \cite{cvetko}.
The $c(4 \times 2)$ symmetry corresponds the ground phase of the Ge(001) 
surface and it was shown to be formed by buckled Ge dimers 
(tilt $\sim 19 ^{\circ}$) with 
an alternate arrangement of the buckling angle 
orientation \cite{ferrer}. 
This phase, mimicking an antiferromagnetic 
ordering of the dimers, undergoes an order-disorder 2D-Ising 
transition to a $(2 \times 1)$ phase at 240~K, thus locally 
preserving the same structure of the low temperature (LT) phase \cite{cvetko}.

Pb is evaporated from both commercial (EPI) and home-made Knudsen 
cells with boron nitride crucibles at a temperature of about 800~K. 
The deposition is monitored in 
real time either by He reflectivity or by X-ray reflectivity, at the 
same time quartz microbalances are used to normalize the deposition 
rate.

In the 110~$^{\circ}$ fixed 
scattering geometry of our HAS apparatus, the perpendicular He momentum transfer is 
$K_{He}^{\perp} = 2 K_{He} \cos 55^{\circ}$, where the He 
wavevector $K_{He}$ can be varied between  6 and 12~\AA$^{-1}$ by 
controlling the temperature of the nozzle source.
Pb deposition was monitored by He reflectivity at fixed He wavevector. 
The step height on the growing surface was 
determined by measuring the He reflectivity as a function of 
the He wavevector after stopping deposition at a given stage. In 
fact, almost no kinetic evolution (recovery) was observed at the considered 
temperature. 
For a stepped surface this measurement gives rise to reflectivity 
maxima and minima due to the interference between the He waves 
scattered by terraces of different height. A rough estimate of 
the step height can be obtained by the 
oscillation period, when only two layers are exposed on the 
surface \cite{lent}.

The X-ray diffraction and reflectivity measurements have been taken at 
the end station of the ALOISA beamline.
X-ray reflectivity has been measured at a fixed grazing angle 
of $\alpha = 8.25$~$^{\circ}$ and for a few photon energies.
In this case, maxima and minima of the reflectivity arise from the 
interference between the X-ray wave reflected by the film/substrate 
interface and that reflected by the film surface. The 
period of the reflectivity oscillation yields the film thickness 
corresponding to the selected perpendicular momentum transfer $K^{\perp} = 2 K 
\sin \alpha$.
Measurements have been performed in a photon energy range from 5 up to 
6.5~keV,   corresponding to periodic interference conditions from film 
thickness ranging from $\Delta D = 8.64$ to 
6.58~\AA, respectively, i.e. from 3 to 2.3 ML, in units of Pb(111) 
bulk interlayer spacings. X-ray reflectivity thus measures the overall film 
thickness as opposed to He scattering, which only probes the 
terrace distribution of the exposed surface layers.
For a few selected deposition stages in proximity of the critical thickness
for the onset 
of layer-by-layer growth, the deposition has been stopped and scans 
with the perpendicular momentum transfer have been taken along the 
$(2,\overline2,0) + L(1,1,1)$ rod 
of the Pb diffraction peak. 
In this case, maxima and minima are 
due to the diffraction grating formed by the interlayer spacings 
of the Pb(111) atomic planes. The perpendicular distribution of the Pb film 
layers can be obtained by fitting a kinematic scattering model to the 
rod scans.

\section{Low temperature layer-by-layer growth}

\begin{figure}
\center
\includegraphics[width=.7\textwidth]{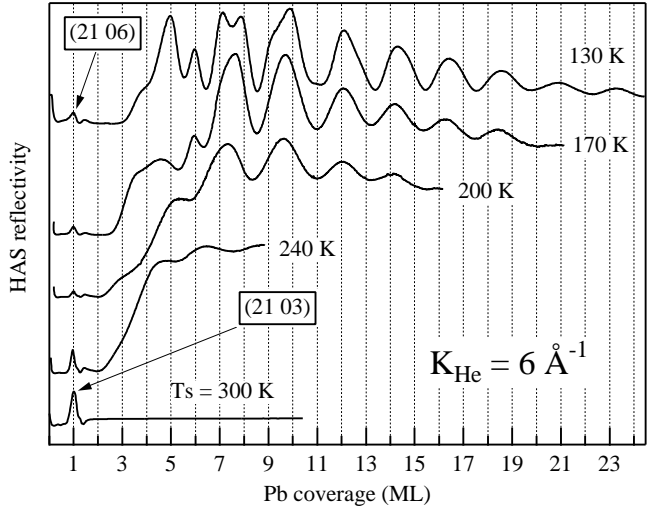}
\caption{He specular reflectivity during Pb deposition taken for a few 
substrate temperatures at He wavevector $K_{He} = 6$~\AA$^{-1}$. 
The scattering plane has been kept 
azimuthally oriented along the 
direction of the main Pb(111) symmetry axis, 
i.e. at 45$^{\circ}$ from the substrate [001] direction.
}
\label{fig1}
\end{figure}

Pb deposition initially leads to the formation of a few ordered 
pseudomorphic structures in the submonolayer range \cite{yang}. The highest 
coverage ordered structure displays a $(21~03)$ symmetry 
reconstruction [which undergoes a transition to a $(21~06)$ symmetry 
phase at LT]
and corresponds to the deposition of about one monolayer (ML) of 
Pb(111), that is about 5/3 ML in units of the Ge(001) atom 
density \cite{bunk}.
The addition of just a few percent of Pb leads to a mixed phase of 
coexisting $(21~03)$ and  $c(4 \times 8)$-incommensurate 
domains \cite{nota1}. Upon further deposition at room temperature (RT) 
large and widely spaced 3D clusters are formed which do not yield a 
coherent contribution to He scattering, thus almost not affecting the 
He reflectivity.
It was previously shown that, at low temperature, 
irregular He reflectivity oscillations can be 
detected with an amplitude increasing as the temperature is 
decreased. The deposition curves for a few 
substrate temperatures are shown in  
Fig.~\ref{fig1}. Layer-by-layer growth conditions are met for substrate 
temperatures lower than $\sim$~130-140~K in the $4 - 12$~ML thickness 
range.\cite{crottini} The irregular shape of the 
oscillations is strongly dependent on the selected He wavevector and 
it is always perfectly reproducible.

\begin{figure}
\center
\includegraphics[width=.6\textwidth]{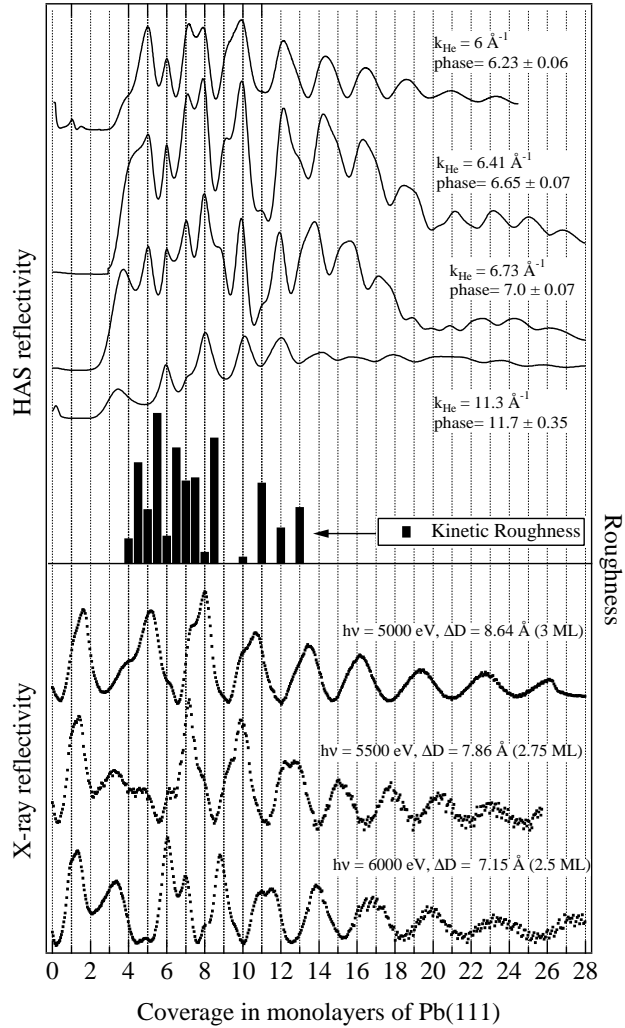}
\caption{Top: He reflectivity curves (full line) 
taken during deposition at T$_{s} = 130$~K 
for four He wavevectors, as indicated by the corresponding labels. 
The scattering plane has been kept azimuthally oriented along the 
$\Gamma M$
direction of the main Pb(111) symmetry axis, i.e. at 45$^{\circ}$ from 
the substrate [001] direction. 
Middle: the vertical bars (right $y$ axis) represent the kinetic roughness, 
i.e. the amplitude of the 
reflectivity oscillations taken as a function of the perpendicular 
momentum transfer, as shown in the left panel of Fig.~3. 
Bottom: X-ray reflectivity curves (filled markers) taken at fixed 
grazing energy for a a few photon energies. The same deposition 
conditions as in the HAS experiments have been used.
}
\label{fig2}
\end{figure}

\begin{figure}
\center
\includegraphics[width=.6\textwidth]{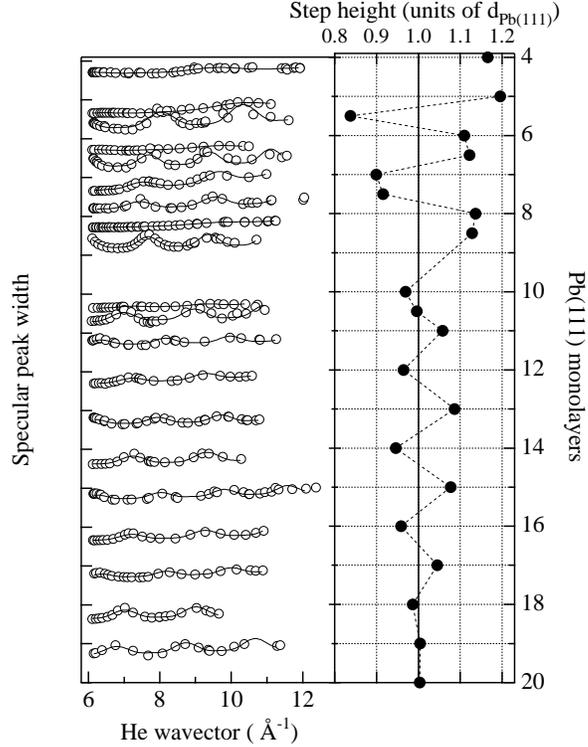}
\caption{Left panel: the width of the HAS specular peak (open circles)
 is taken as a 
function of the He wavector for different Pb coverage. 
Each curve has been simulated (full fitting line) 
with a two-level model  
by using the coverage of the two levels and the step height as fitting 
parameters. The curves have 
been vertically shifted by a constant offset proportional 
to the Pb coverage, which is reported on the right $y$ axis.  
Right panel: the step height as determined from the analysis of the left 
panel is reported as a function of the Pb coverage, 
after Ref.~\cite{crottini}. The 
step height (horizontal top axis) is reported in units of the 
Pb(111) interlayer spacing. 
}
\label{fig3}
\end{figure}

A set of He reflectivity curves taken at different He wavevectors is 
shown in Fig.~\ref{fig2} for deposition on the substrate held at T$_{s} = 
120-130$~K. All the reflectivity curves have been normalized to a 
constant deposition rate by quartz microbalance measurements. The 
deposition rate has been calibrated a posteriori by a detailed 
analysis of the reflectivity as a function  of the perpendicular momentum 
transfer $K_{He}^{\perp}$ for
each deposition stage (see left panel in Fig.~\ref{fig3}) \cite{nota2}.
The corresponding oscillation of the (0,0) peak width and intensity 
yields the height of the terrace step \cite{lent}. 
Due to the high interference 
order, the step height can be 
accurately determined by the positions of maxima and minima.
At the same time, the amplitude of the oscillations is proportional to the 
degree of surface roughness (density of steps).
In this way, a consecutive sequence of alternating flat and rough surface 
has been identified in the first 7-8 oscillations (see the right 
panel in Fig.~\ref{fig3}), 
corresponding to a layer-by-layer regime of growth \cite{crottini}.  

Maxima and minima corresponding to the deposition of each new layer 
are easily detected in the curves of Fig.~\ref{fig2}. 
Interestingly, the roughness minima detected in the layer-by-layer 
regime, which correspond to 
extremely flat surfaces, occur at the completion of even-layers 
deposition, i.e. at 4, 6, 8 and 10~ML. Less flat surfaces are observed 
in correspondence with odd-layers completion, thus indicating that Pb 
films of even number of (111) Pb layers are energetically more stable. 
Interestingly, Tringides et al. found that the most favored Pb
islands on Si(111) have odd-layer heights (5, 7 and 9) 
\cite{hupalo1,hupalo2,yeh}. However their 
Pb island heights 
were measured by LEED and STM relatively to the pseudomorphic 
layer \cite{hupalo2} 
and not to the substrate surface, as in the present case. 

The highest sensitivity to the surface roughness 
is obtained at $K_{He} = 6.73$~\AA, which corresponds 
exactly to an out-of-phase scattering condition for the Pb(111) bulk 
interlayer spacing. Only the flattest stages yield intensity maxima 
at  $K_{He} = 11.3$~\AA, partly for its close in-phase scattering 
condition, mostly because of the larger spread in the He wavector at 
this energy, which smears the interference among the He waves 
scattered by different terraces. 
In this case, the oscillations mainly display a bi-layer periodicity. 
Interestingly, also the other curves display a bi-layer modulation of 
the layer-by-layer oscillations, which evolves towards bi-layer 
oscillations beyond 10-12~ML.
We observed a similar behavior also for the LT deposition of Pb on 
Ge(111), where the main difference appears to be that lower substrate 
temperatures are required to have a complete substrate wetting \cite{pbge111}.
Double layer growth was also observed by HAS  to be favored  at certain 
thickness for LT deposition of Pb 
on Cu(111) \cite{hinch} and a bi-layer periodicity of the HAS 
reflectivity was observed for Pb on Cu(100) \cite{zeng} and 
Si(111) \cite{schmicker}.
According to the energy balance model of Zhenyu Zhang, this 
bi-layer periodicity is due to the Friedel's oscillation of the Pb 
electron density in a semi-infinite slab. 
In fact, consecutive layers correspond to alternating 
maxima and minima of the electron density perpendicular profile, the 
minima favoring the stabilization of a flat surface \cite{zhang}.

\begin{figure}
\center
\includegraphics[width=.9\textwidth]{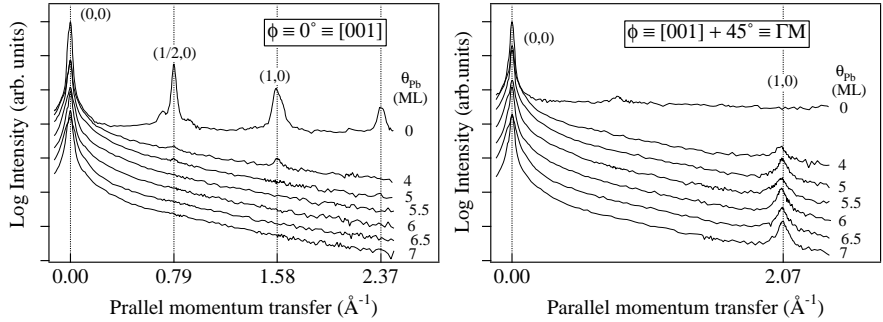}
\caption{Left panel: He diffraction scans taken along the [001] direction of 
Ge for different coverage of Pb. The half-integer peaks 
of the clean $c(4 \times 2)$ Ge(001) surface are also present in the 
mixed $(21~06)$ + $c(4 \times 8)-i$ phase.
Right panel: He diffraction scans taken along the $\Gamma M$ Pb(111) 
direction, i.e. at an azimuthal angle $\phi = 
45^{\circ}$ 
from the [001] direction of Ge, for the same Pb coverage of the left 
panel. The diffraction peaks of the Pb(111) surface are only observed.
}
\label{fig4}
\end{figure}

\subsection{Wetting of the low temperature film}

It must be pointed out that, in order to observe the interference 
effect between the growing layers, Pb growth must proceed 
coherently over a large surface area, since HAS scattering is averaged 
over an effective illuminated area of about 0.5~mm$^{2}$. In the 
early stage of deposition, growth seems to proceed via random 
nucleation of 3D Pb clusters and
the HAS diffraction peaks of the Pb(111) surface 
are first observed at about 4~ML.  
The Pb(111) surface is incommensurate with the substrate and its main 
symmetry axis $\Gamma M$ is found to be oriented at $45^{\circ}$ with respect to 
the [001] substrate direction. In fact, two Pb(111) domains 
rotated by $90^{\circ}$ are found, which is the only fingerprint left from the 
substrate squared lattice.  

HAS diffraction from a Pb(111) surface first appears at $\theta = 4$~ML.
At this coverage, the Pb film is not yet wetting the 
whole substrate, in fact the diffraction peaks originated by the 
substrate pseudomorphic structure are still detected (see left panel 
of Fig.~\ref{fig4}). This could be due either to fragmentation of the 
film into flat islands, as commonly observed for Pb on Si(111) 
\cite{hupalo1,su}, or to the occurrence of deep pits uncovering the 
substrate, as observed for Ag on Si(111) and 
GaAs(110) \cite{huang,smith}. However, just after the deposition of one 
additional 
layer, the substrate peaks disappear and the Pb(1,0) peaks reach 
their nominal position \cite{nota3} without any further shift as the thickness 
increases, i.e. no lateral strain is observed beyond 5~ML.  
At this coverage, we cannot exclude the 
occurrence of small pits, which would be hidden due to shadowing 
effects. At 5~ML of Pb, the pits uncovering the pseudomorphic Pb layer
would be 4-layer deep 
and their maximum lateral size for not being detected would be 3~nm, 
to be compared with an average domain size of 13~nm (calculated from 
the Pb(111) peak width) or an average terrace size of 25~nm 
(calculated from the specular peak width). 

X-ray reflectivity 
measurements, shown in the bottom of Fig.~\ref{fig2}, also support 
the model in which 
a flat surface is formed at the completion of the 5th Pb monolayer. 
The X-ray reflectivity oscillations have a thickness 
periodicity corresponding to the selected momentum transfer, but, in 
case of a perfect layer-by-layer growth, cusp-like features are 
superimposed to the oscillations, due to an overall increase of the 
reflectivity whenever a flat surface is formed \cite{weschke}. 
We have fitted the X-ray reflectivity to a simple growth model with 
the scattered intensity  calculated as:
\begin{equation}
	I(t) = \left| F_{Ge} + \sum_{1}^{n} \theta_{n}(t) \cdot f_{Pb} \cdot 
	\rho_{Pb} \cdot e^{i 
	K^{\perp} z_{n}}\right|^{2} ;
	\label{eq1}
\end{equation}
where $F_{Ge}$ and $f_{Pb}$ are the Ge substrate structure factor 
\cite{ferrer} and Pb
atomic scattering factor \cite{henke}, respectively, $\rho_{Pb}$ is 
the atom density of a Pb(111) layer,
$z_{n}$ is the perpendicular coordinate of the $n$-layer and 
$\theta_{n}(t)$ is the instantaneous coverage of the $n$-layer. The 
layer filling as been described according to a conventional birth-death 
model \cite{cohen}:
\begin{equation}
\frac{d\theta_{n}}{dt} = (1-\eta_{n-1}) \frac{R}{\rho_{Pb}} 
(\theta_{n-1}-\theta_{n}) + \eta_{n} \frac{R}{\rho_{Pb}} 
(\theta_{n}-\theta_{n+1}) ;
\label{eq2}
\end{equation}
where $R$ is the deposition rate (in atom per second and unit area) 
and the interlayer diffusion 
probability $\eta_{n}$ is a phenomenological effective parameter 
describing both the intra- and inter-layer atom diffusion \cite{vandervegt}.
All the interlayer Pb(111) spacings have been kept fixed to the bulk 
value, while the separation $d_{Pb-Ge}$ between the first (bottom) 
Pb layer and the 
substrate has been fitted to obtain the interface thickness. This 
parameter strongly affects the average reflectivity intensity, which 
is asymptotically matched at high coverage.
 
 \begin{figure}
 \center
\includegraphics[width=.7\textwidth]{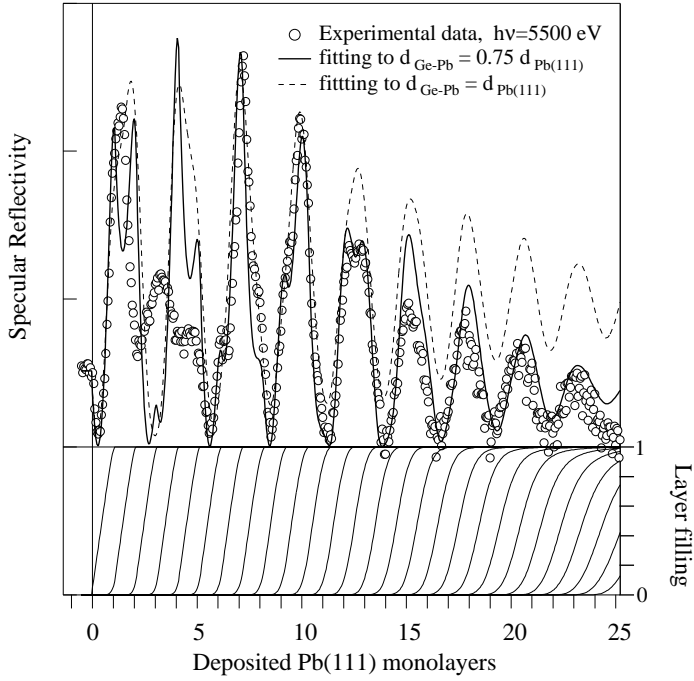}
\caption{Top: the X-ray reflectivity taken at 5500~eV during Pb 
deposition at 130~K (open circles) is simulated by a birth-death 
growth model. Layer-by-layer growth is established from the 5th Pb ML.
The best fit yields a contracted interface thickness 
(full line), whereas the average intensity of the reflectivity is 
overestimated when using the unrelaxed Pb(111) spacing for the 
interface width (dashed curve). 
Bottom: the single layer filling is reported on the right axis.
}
\label{fig5}
\end{figure}

As can be seen in 
Fig.~\ref{fig5}, a contracted interface 
thickness has been found corresponding to $d_{Pb-Ge} = 0.75 d_{Pb(111)}$.
From Fig.~\ref{fig5}, it can be seen that the first cusp-like 
feature, corresponding to a flat surface, is found after the 
deposition of 5~ML, a similar feature is observed at any photon 
energy (compare with X-ray reflectivities shown in Fig.~\ref{fig2}). 
At lower coverage, the experimental data 
show similar oscillations for all the considered perpendicular momentum 
transfer, which cannot be reproduced by any layer-by-layer growth 
simulation. From comparison with HAS reflectivity, we can say that in 
this coverage range the Pb film is growing with 3D islands, which are 
also undergoing a morphological transformation towards an homogeneous 
flat film of 5~ML.

\subsection{Metastability of the low temperature film}

The Pb film, grown layer by layer below 130~K, is always found to be 
kinetically frozen  for thickness up to a few dozen  monolayers.
Whenever the LT Pb film is annealed to the room temperature, the Pb(111) 
diffractions peaks disappear and only the diffraction from the 
pseudomorphic phase is detected (both with HAS and X-ray 
scattering), thus suggesting a fragmentation of the film 
into large clusters 
uncovering the Pb/Ge interface.

\begin{figure}
\center
\includegraphics[width=1.0\textwidth]{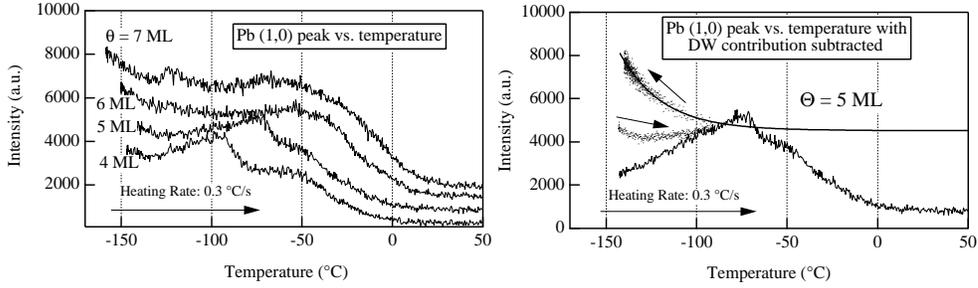}
\caption{Left panel: the (1,0) He diffracted intensity of the Pb(1111) 
sample 
is shown as a function of the temperature 
at a constant annealing rate.
Four scans are shown, corresponding to the measurements taken on Pb films 
of different coverage. The diffraction from the Pb(111) surface 
irreversibly disappears above $\sim$~-50$^{\circ}$C.
Right panel: by annealing below the deconstruction temperature (dotted 
curve, arrow pointing right) the Pb(111) surface quality (domain 
size) is increased as witnessed by the strong intensity increase 
obtained upon cooling down (dotted curve, arrow pointing left). The 
Debye-Waller exponential attenuation (full line) 
on the given Pb film can be obtained 
from the cooling branch of the temperature ramp. The full heating 
ramp displayed in the left panel at 5~ML is also shown for 
comparison after the Debye-Waller 
attenuation correction.}
\label{fig6}
\end{figure}

In fact, the LT smooth and wetting film appears to be metastable, 
as substrate temperatures lower than 130~K are required for the 
onset of regular layer-by-layer growth, but the surface quality 
(terrace and domain size as obtained by the diffracted peak widths) 
can be increased by short annealing up to 190~K of the Pb film after 
deposition. The intensity of the Pb(1,0) diffracted peak, shown in 
Fig.~\ref{fig6}, has been 
taken for a few film thickness during annealing at a constant rate.
The Pb(1,0) irreversibly disappears beyond a deconstruction temperature 
ranging from 190 to 220~K, depending on the film thickness. 
If the film is annealed below the critical temperature and then cooled 
down, the Pb(111) domain size can be almost doubled. 
Further deposition on the annealed film yields reflectivity 
oscillations with larger amplitude (for both HAS and X-ray 
reflectivity), i.e. more perfect layer-by-layer growth conditions are met.
Similarly, pre-deposition at RT of 1~ML, leading to the formation of 
a pseudomorphic $(21~03)$ phase with larger domains, yields better 
layer-by-layer growth upon further deposition at LT.

\begin{figure}
\center
\includegraphics[width=.7\textwidth]{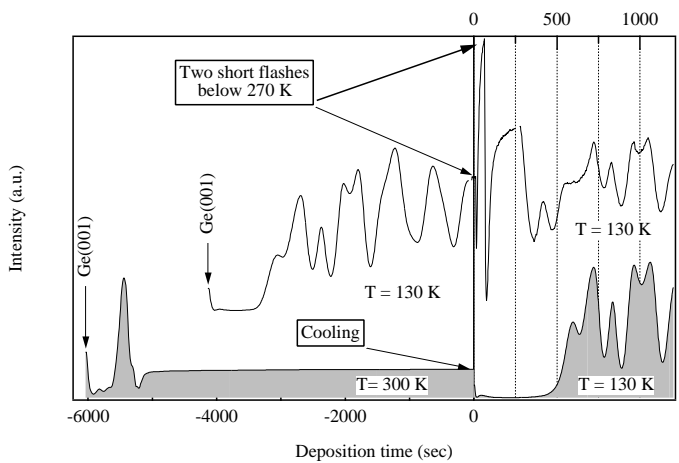}
\caption{Upper curve:  He 
reflectivity during Pb deposition up to 13~ML at 130 K, then the Pb 
shutter is closed (t=0). A short flash at 200~K yields a better ordered 
Pb(111) film. A 
second flash at 250~K causes the fragmentation of the Pb film.
 Further deposition (t=270) at 130~K yields 
the layer-by-layer growth features characteristic 
of the deposition on the clean substrate. 
Lower curve (shadowed): 
extended Pb deposition at 300~K does not yield any 
growth feature after the formation of the pseudomorphic phase at 
1~ML. 
The deposition is stopped and the sample is cooled down to 130~K, then 
Pb deposition is started again (t=0) with the same rate. The typical LT 
layer-by-layer growth oscillations set in again.
All curves have been taken with the same He wavector of 6~\AA$^{-1}$ 
and along the same azimuthal orientation [001]+45$^{\circ}$.
}
\label{fig7}
\end{figure}

The fragmentation of the wetting Pb film can be demonstrated by the 
measurement shown in Fig.~\ref{fig7}, where 13~ML of Pb are deposited  
at LT. First the film is well ordered by a short flash to 200~K, then the 
decomposition is activated by a flash to 250~K and quickly frozen 
by a rapid quench to 130~K. Further LT deposition yields the same 
layer-by-layer growth oscillations, 
as for original deposition on the clean substrate.
We must conclude that Pb growth on existing large 3D islands yields an 
uncorrelated contribute to the He scattering, while the oscillations 
are due to coherent growth on the uncovered Pb/Ge interface. The 
reduced amplitude of the reflectivity oscillations reflects the 
smaller uncovered area of the substrate, where coherent layer-by-layer 
growth can take place.

A similar behavior is observed when depositing Pb at RT, where no 
ordered structure formation is observed after 
the formation of the pseudomorphic phase at 
1~ML, and then cooling down below 130~K to continue the deposition (as 
shown in Fig.~\ref{fig7}): 
layer-by-layer growth is observed again, as for a clean 
substrate. The only difference with the reflectivity measurements of 
Fig.~\ref{fig2} is the time required for the onset of layer-by-layer 
growth, which is delayed on the pre-deposited substrate. The Pb  
islands thus act as a sink for the diffusing adatoms, delaying the 
formation of the flat film.

\begin{figure}
\center
\includegraphics[width=.7\textwidth]{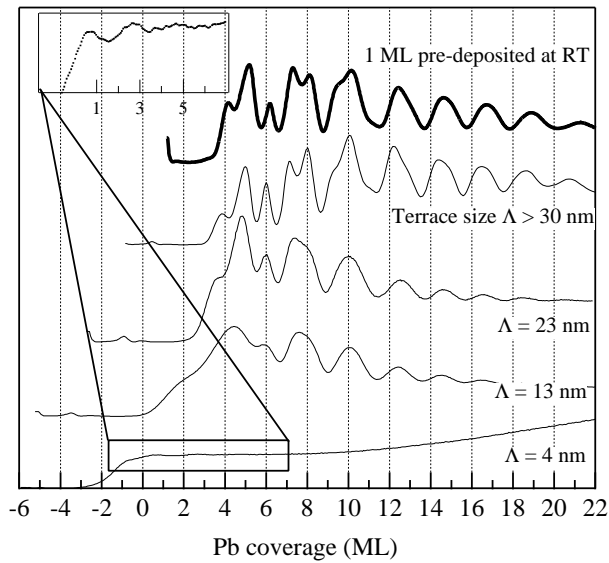}
\caption{He reflectivity taken during LT Pb deposition on Ge surfaces
 with different average terrace size $\Lambda$. Each curve is labelled by the 
 corresponding Ge(001) terrace size (in nm). The top curve (thick
 line) has been 
 obtained by deposition of the 1st Pb ML at RT and further deposition 
 at LT. }
\label{fig8}
\end{figure}

The sharpness of the interface seems to play a fundamental 
role in the onset of layer-by-layer growth. 
Pb deposition at 130~K has been performed on Ge(001) 
of different average terrace size, as obtained by controlled ion 
erosion \cite{cvetko1}. As can be seen from Fig.~\ref{fig8}, the larger 
the terrace 
size is, the  larger are the amplitudes of the reflectivity 
oscillations. As the average terrace size is decreased one observes 
an increasing delay in the onset of layer-by-layer oscillations and 
an anticipation of the bi-layer modulation. 
For the narrower terrace width, only faint 
double layer modulations are observed, which are superimposed on a 
rather intense He reflectivity background indicating the formation of 
a metallic surface.

\section{Surface relaxation of the growing film}

\begin{figure}
\center
\includegraphics[width=.9\textwidth]{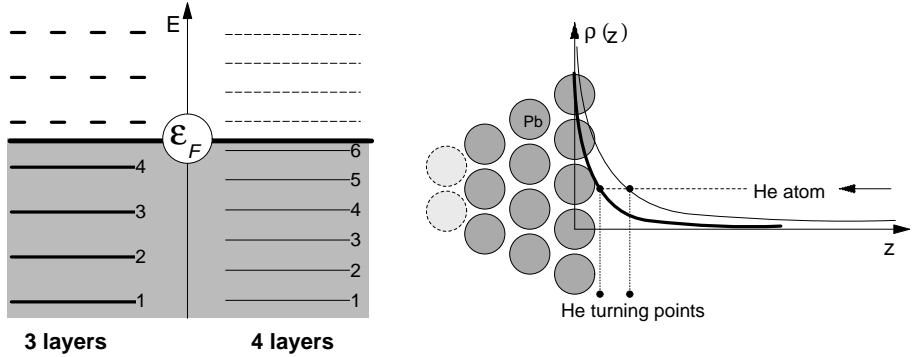}
\caption{Left panel: the individual bound states in a 3-layer thick 
Pb film are indicated with full thick lines at the left-hand side; four states 
are accommodated below the Fermi level. The thickness increase by one 
layer allows two new bound states to fall below the Fermi level
(thin full lines on the right-hand side).
Right panel: the in-vacuum tail decay of the surface charge density 
$\rho (z)$ is schematically represented for the two cases 
of the left panel. Thick and thin lines are 
assigned to the 3- and 4-layer thickness Pb films, respectively.
The two iso-density perpendicular coordinates for the He turning point 
(and for the STM tip distance from the surface) are also shown.
}
\label{fig9}
\end{figure}

Concerning the irregular, but reproducible, 
shape of the HAS layer-by-layer oscillations,
a detailed study of the reflectivity as a function of $K_{He}$ found 
it to be originated by an oscillatory variation of the apparent monatomic 
step height as new layers are added (see right panel of 
Fig.~\ref{fig3}), 
i.e. to an oscillatory change of the interference 
conditions from the Pb(111) terraces \cite{crottini}.
These oscillations have been attributed to QSE, 
i.e. to perpendicular displacements of the surface charge density 
occurring whenever new QWS can be accomodated below the Fermi level 
as the film thickness $D_{Pb}$ is increased.
This mechanism has been experimentally confirmed by STM spectroscopy 
on individual flat Pb islands grown on Si(111), where the quantum 
numbers have been assigned to each quantized state for islands of 
different thickness \cite{su}. It has been shown that adding a new 
layer increases the number of bound states below the Fermi level by 
either one or two for islands of consecutive layer thickness
(recall that three new bound states can be accommodated for each 
bi-layer thickness increase, since $d_{Pb} \sim \frac{3}{4} 
\lambda_{F}$), which corresponds to a cyclic variation of the energy 
separation between the Fermi level and the highest 
occupied state (see left panel of Fig.~\ref{fig9}).
As a consequence the slope of the in-vacuum tail decay of the total 
surface charge 
density $\rho (z)$ is also changing, and the corresponding turning 
point for He atoms scattered by the surface is perpendicularly displaced 
in an oscillatory fashion (see right panel of Fig.~\ref{fig9}).

In fact, in a critical revision of the HAS study published in 
Ref.~\cite{hinch}, Toennies et al. argued that the apparent oscillations of 
the step height are to be entirely attributed to the cyclic displacement of 
the He turning point due to the electronic QSE \cite{braun}, while ab 
initio calculations predict the ionic cores to follow the 
displacement of the charge density, even if to a lower extent
\cite{feibelman,batra}.
On the other hand, the phenomenological model used in Ref.~\cite{hinch}
 to estimate the in-vacuum 
charge spillout was supported by ab initio 
calculation for a Pb(111) slab of up to 15 layers \cite{materzanini}. 
Slight oscillations of the surface relaxation ($\pm 1$\%) and of the topmost 
interlayer spacings were predicted as a function of the slab 
thickness, which did not alter the average -5\% compression of the topmost 
to second layer spacing and the average +2\% expansion of the 2nd to 3rd 
layer spacing. These calculated average values are 
in good agreement with the experimental ones obtained 
on a Pb(111) bulk sample \cite{li}.

\begin{figure}
\center
\includegraphics[width=.6\textwidth]{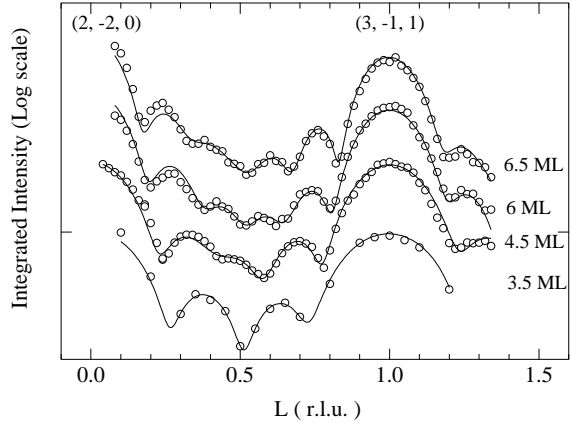}
\caption{X-ray diffraction scans of the (2,$\overline2$,0)+L(1,1,1) rod 
taken after the deposition of 3.5, 4.5, 6 and 6.5~ML of Pb. The 
experimental data (open circles with error bars) are simulated with a 
structural model as described in the text (full line). 
}
\label{fig10}
\end{figure}

In order to discriminate between the different relaxation models, 
we used surface X-ray diffraction to measure the 
perpendicular distribution of the Pb(111) layers within the growing film.
For a few selected thicknesses we collected the X-ray data as a function
of the perpendicular momentum transfer along the Pb(111) crystal truncation
rod, i.e. the diffracted intensity distributed in the reciprocal space among
the Bragg reflections. The Pb(111) 
layer spacings can be determined by fitting the 
X-ray interference fringes detected along the rod scan. 
In particular, we measured the 
(2,$\overline2$,0)+L(1,1,1) rod from the in-plane (2,$\overline2$,0) 
peak (L=0) up to L$\sim$1.4, through the out of plane 
(3,$\overline1$,1) peak (L=1). Each point along the rod scan has been 
individually fitted in order to obtain the appropriate background 
subtraction and integrated intensity, as reported in Fig.~\ref{fig10}. 

First of all, it must be pointed out that the Pb(111) in-plane 
diffraction peak first appears at a Pb coverage of $\sim$~2.5~ML, in 
contrast to the amorphous growth reported for Pb on Si(111), 
where the XRD peaks of crystalline Pb(111) are first 
observed after 5 Pb monolayers 
\cite{edwards}. 
At the coverage of 2.5~ML, the diffracted intensity is not high enough 
for measuring the entire rod scan, 
however at least three Pb(111) layers can be observed 
approaching the coverage of 3~ML. 
Keeping in mind that a large uncovered area of pseudomorphic 
layer is clearly detected by HAS at this coverage, 
we can conclude that the X-ray diffraction yield is 
given by largely spaced 3D islands.

We did not directly measure the structure evolution of the 
pseudomorphic layer
 below the growing Pb(111) islands.
In fact, the stability of a pseudomorphic phase was observed by 
X-ray diffraction for Pb on Si(111), where the diffraction peaks from 
the Pb-induced $(7 \times 7)$ reconstruction were still detected after 
the deposition of 1800~\AA~ of Pb \cite{howes}. 
However, the same 
authors found that the  $(\sqrt3 \times \sqrt3)-R30^{\circ}$ 
Pb-induced reconstruction is destroyed by the growing islands 
\cite{howes}. The different number of metal atoms per unit cell, which are 
covalently bound to the substrate, probably determines the 
stability criteria of the pseudomorphic layer.
In our case, when entering in the layer-by-layer growth regime, 
the number of Pb(111) layers 
is found to be equal to the number of deposited monolayers, thus 
suggesting the deconstruction of the pseudomorphic layer to be 
incorporated in
the Pb(111) film (recall that the pseudomorphic layer as practically 
the same atom density of a Pb(111) layer \cite{bunk}).
Both the appearance of the first Pb islands with a three-layer 
thickness and the onset of layer-by-layer growth at a 
thickness of five Pb(111) layers qualitatively agree with the 
magic and critical thickness predicted by the Zhang's model for Pb 
growth on semiconductors \cite{zhang}. However, the enhanced surface 
flatness observed by HAS at even-layer multiples (see the low kinetic 
roughness at 4, 6, 8, 10~ML in the middle panel of Fig.~\ref{fig2}) 
asks for more refined 
calculations to understand the role played by the pseudomorphic 
interface.

In  a kinematical scattering scheme, the integrated intensities 
of diffraction peaks are proportional to the modulus square of 
the crystal structure factors, whose magnitude can be simulated 
in order to retrieve the atomic plane positions. The rod scan 
simulations 
were computed using both home-made routines \cite{brunof} 
and the freely distributed ROD program \cite{vlieg}, 
which yielded equivalent results. 
Since the Pb(111) layers are incommensurate with the substrate 
and no accidental superposition with the Ge rods occurs, 
the calculation can be performed for a free standing Pb film.  
 The interlayer spacings, the 
Debye-Waller attenuation factors and the layer fillings have been used 
as fitting parameters. For the 6 and 6.5~ML Pb coverage films, 
we have reduced the total number 
of fitting parameters (up to 18 could be considered) 
by considering a complete 
layer filling except for the topmost layer, as suggested by the HAS 
and X-ray reflectivity measurements. 
A good reproduction of the measured curves was possible only 
by assuming a strong layer-dependence of the Debye-Waller (DW) factors. 
The DW factor for a bulk Pb crystal is available from 
the literature  
as derived from experimentally determined phonon density of states 
\cite{pend}. 
For a crystal temperature of 130~K, the thermally averaged rms displacement 
of a bulk atom from its equilibrium position is expected to be 
$\sqrt{<u^{2}>} = 
0.11$~\AA. 
In general we found larger DW factors, corresponding to displacements 
spanning from $\sim$~0.15~\AA,
for the deeper layer, up to $\sim$~0.45~\AA~ for the topmost one. 
While a larger average displacement may be expected for a surface 
atom than for a bulk one, other mechanisms could be involved. 
In fact, large terraces are 
formed during LT layer-by-layer growth, which do not evolve
after interruption of the evaporation. This implies a very  high 
mobility of the Pb surface atoms, which contributes to the
increase of the rms displacement.

\begin{figure}
\center
\includegraphics[width=.6\textwidth]{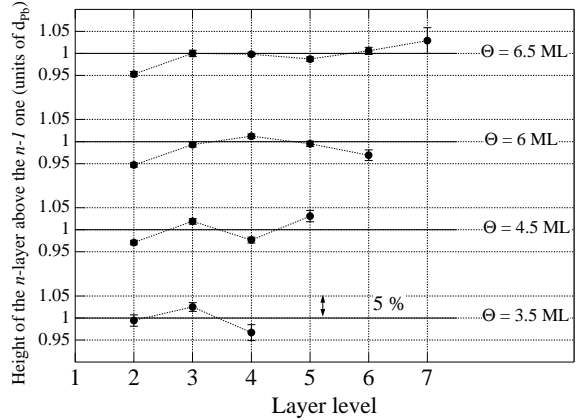}
\caption{Interlayer spacings in units of $d_{Pb(111)}$, 
as obtained by fitting to the rod scans 
of Fig.~10. For each film the height of the $n$-layer above the 
$(n-1)$-layer is given. 
The 1st layer distance from the interface 
cannot be determined since the simulation has been performed for a 
free-standing Pb(111) film.
}
\label{fig11}
\end{figure}

The hierarchy of the 
layer depth is given both by the finite value of the X-ray penetration 
length for Pb (20~\AA), yielding different weights for different
layers, and by the hierarchy of the DW factors (the smaller the 
deeper). 
As can be seen from Fig.~\ref{fig11}, the distance between the first 
(bottom) and second layer is always contracted up to $\sim 5$~\% of 
the bulk Pb(111) interlayer spacing. Given the free 
standing slab model used for the simulation, no information can be 
obtained about the height of the bottom layer of Pb(111) above 
the interface (whose thickness was also found by X-ray reflectivity to be 
contracted).
More interestingly, it is clearly seen that the topmost layer height 
oscillates around the bulk value as new layers are added. In contrast 
to the calculations by Materzanini et al. \cite{materzanini}, 
the atomic plane of the 
topmost layer is effectively following the surface charge density in 
its oscillatory behavior, yielding alternative expansion/compression 
of a few percent, as predicted by early calculations \cite{feibelman,batra}.
As expected \cite{batra}, also the inner layer spacings display 
small relaxations (within 1-2~\% ). 
However, given the error bars and the increasing number of fitting 
parameters as the thickness increases, it is not possible to single 
out a general behavior for the inner layers. 
Only the subsurface layer seems to display 
an oscillating behavior 
(opposed to the oscillation of the topmost layer relaxation)

\begin{figure}
\center
\includegraphics[width=.6\textwidth]{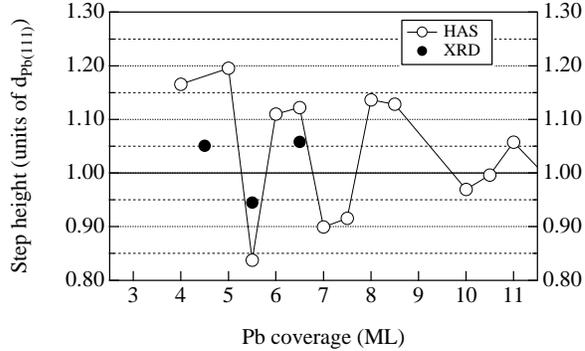}
\caption{The apparent step height obtained by direct HAS measurement 
is shown 
(open circles) together with the step height obtained by subtracting 
the XRD topmost layer spacings of consecutive film thickness (filled 
circles). The step height is given in units of $d_{Pb(111)}$.
}
\label{fig12}
\end{figure}

The absolute amount of the topmost layer relaxation is much lower 
than the amplitude of the apparent step height oscillation, as 
measured by HAS. In fact, a direct comparison between these two 
quantities is misleading, since the XRD analysis yields the 
height of a layer on top of another one, while HAS measures the 
apparent step height separating two adjacent terraces, i.e. two 
topmost layers of different height level. A 
more suitable term of comparison is the height difference between the 
XRD topmost layer of two consecutive thickness Pb films.  The XRD step 
height is thus evaluated as $h_{n}^{XRD} = 
(d_{n}^{ top} - d_{n-1}^{ top}) + d_{Pb(111)}$ and shown in 
Fig.~\ref{fig12} 
in comparison with the HAS apparent step height.
In this case the variation of the step height becomes relevant 
(exceeding $\pm 5$~\% of the bulk Pb(111) layer spacing), even if 
it is still lower than the corresponding value obtained by HAS. Most 
importantly the step height oscillations effectively follow the 
surface charge density oscillations measured by HAS. 
The surface relaxation is thus 
strongly affected by QSE.  A similar conclusion was 
indirectly obtained by the detailed STM spectroscopy 
study of Tsong and coworkers \cite{su}, who measured the oscillations 
of the apparent step height as a function of the tip bias voltage
and found them to be present also for a reversed bias, thus 
indicating an effective structural variation of the step height.

\section{Surface vibrations of thin Pb films}

\begin{figure}
\center
\includegraphics[width=.8\textwidth]{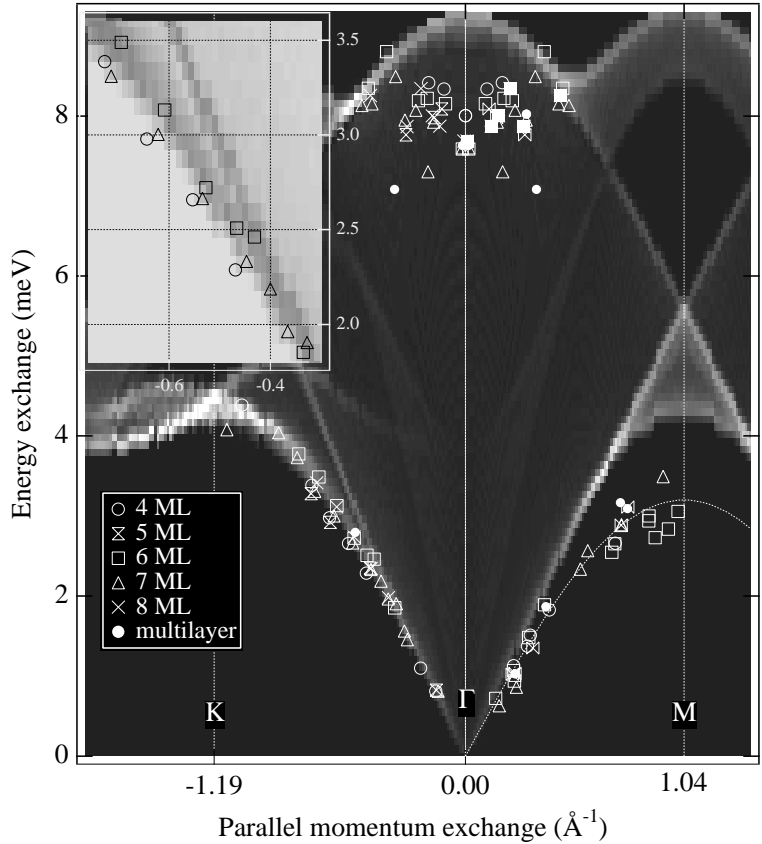}
\caption{The phonon spectra measured for a few thin Pb films of 
different thickness (open markers) and one thick film (filled marker) 
are shown in the first Surface Brillouin Zone (SBZ). 
The bulk vibrations, projected 
on the Pb(111) surface are also shown as shadowed bands, whose 
brightness is proportional to the corresponding cross-section for He 
inelastic scattering. In the inset (reversed gray scale), a small 
portion of the SBZ is enlarged to show the difference among phonon 
energies of 4, 6 and 7~ML Pb films. 
}
\label{fig13}
\end{figure}

The variation of the surface relaxation due to QSE, although small, 
is sizeable and led us to study its influence
on the Pb film vibrational properties by inelastic He scattering.  
If the  variations of the topmost interlayer distance observed for different 
film thicknesses were large enough, they may be expected to alter also the 
force constants of 
the growing film and therefore modify the acoustic vibrations of the film. 
The phonon dispersion curves of the Pb(111) vibrational modes have been
measured along the $\Gamma M$ and $\Gamma K$ (at 30$^{\circ}$ from $\Gamma M$) 
high symmetry directions 
within the 1st Brillouin zone (BZ) of the Pb(111) surface and compared for 
different Pb film thicknesses where the largest QSEs have been observed.
The Time of Flight (TOF) spectra of low energy He atoms 
($E_{He}=19$ meV) have been acquired 
for different scattering geometries.
All measurements have been performed at 130~K, where Pb thin films 
are stable enough to allow long time phonon data acquisitions. 
From the position of the inelastic peaks in the TOF spectra, 
the scattering events in which He atoms have exchanged energy and momentum 
quanta at the surface may be identified, 
i.e. scattering events in which surface phonons 
are created/annihilated. The resulting phonon dispersion obtained 
from over 20 TOF spectra taken at different incidence angles 
is shown in Fig.~\ref{fig13}.  Since there are  
two 90$^{\circ}$-rotated domains 
of Pb(111),  both $\Gamma M$ 
and $\Gamma K$ phonon dispersion curves are detected along the same 
surface direction.  
Two distinct acoustic branches have been identified, one with a flat 
dispersionless behavior at $1.04$~\AA$^{-1}$ , corresponding to the 
Brillouin zone boundary along the $\Gamma M$ surface direction and
 another one at higher frequencies, dispersing up to 1.19~\AA$^{-1}$  
($K$ point 
along the $\Gamma KM$ direction). Both branches have been 
identified a posteriori as acoustic branches along 
the $\Gamma M$ and $\Gamma KM$ directions, respectively.  
This discrimination was not possible for high energy points 
appearing as an optical phonon branch close to the $\Gamma $ point, 
thus they are reported along both $\Gamma M$ and $\Gamma K$ directions.

For comparison, Fig.~\ref{fig13} also shows the calculated vibrational 
spectra of an 
infinite  Pb crystal projected on the (111) direction with Pb force 
constants obtained from Ref.~\cite{landolt}. The shadowed area is obtained as  
the density of the bulk projected vibrational modes weighted with 
the component of the mode polarization along the surface  normal 
in order to mimic the He inelastic cross section \cite{nardelli}. 
The experimental points along both $\Gamma M$ and $\Gamma K$ directions 
fall below the bulk projected area, confirming the surface character 
of the detected vibrational modes. 
An appreciable difference among the phonon energies of different 
films has been detected over a wide range  of momentum exchange 
(see the inset panel in Fig.~\ref{fig13}). For instance, along the
$\Gamma K$ surface direction, the observed phonon frequencies of the 6~ML Pb 
film  are found to be higher than those of the 7~ML Pb film 
($\sim $~5~\%). However, a clear thickness dependent variation of the phonon 
frequencies could not be assigned.
In fact, the effect of force constants variation due to 
topmost layer stretching/contraction is smeared or  even 
cancelled out by the inner layer contraction/stretching.

\section{Conclusions}

Pb deposition on Ge(001) has been shown to proceed layer-by-layer 
when the substrate is held at temperatures lower than $\sim$~130~K. 
Similar findings have been found by a few authors for Pb deposited 
on Si(111) at even lower temperatures 
\cite{jalochowski3,schmicker,edwards}. Such 2D heteroepitaxial growth 
may be expected from
energy balance models when the Pb films exceed a critical 
thickness of five layers \cite{zhang}. In fact, the occurrence of an initial 
transient regime, where Pb deposition proceeds by island growth, is 
commonly observed. The observed critical 
thickness seems to depend not only on the chemical nature of the 
substrate \cite{chao}, but also on its temperature and terrace size. 
X-ray reflectivity  appears to be the most reliable technique to 
measure the critical thickness, since it
effectively probes the whole film. Both our 
XRR measurements and those for Pb/Si(111) \cite{edwards} 
indicate the formation of a 
flat Pb surface after the deposition of 5~ML. At this thickness, 
a 2D growth is observed by both RHEED \cite{schmicker} and HAS 
\cite{schmicker,crottini} 
surface, yielding characteristic layer-by-layer oscillations. 
However, the transition from the island growth to the smooth growth
might not be abrupt. In fact, our HAS data indicate that already 
at 4~ML a flat morphology is formed (see Fig.~\ref{fig2}), but a 
complete wetting of the substrate is not met up to 5-6~ML. 
The complete wetting of the substrate could be driven by 
the mechanism envisaged for Ag deposited on Si(111) 
\cite{huang} and GaAs(110) \cite{evans}, where a 
film with a flat surface of six layers is already formed at 
a coverage lower than 
6~ML with several empty pits having steep perpendicular walls 
reaching the interface. Further Ag deposition leads to the formation of 
the next layers and pits filling.

For Pb on Ge(001), the LT growth could be described as follows.
The first monolayer forms a pseudomorphic layer 
having almost the same atomic density of the close packed (111) plane 
\cite{bunk}, 
then Pb clusters are formed, which soon display a Pb(111) crystal 
structure (in contrast to Pb on Si(111), where the crystal structure 
is formed at 5~ML). Flat Pb islands, probably interconnected, 
of homogeneous height level are formed 
at 4~ML, as witnessed by the appearance of the first maximum in HAS 
reflectivity.  A complete substrate wetting is achieved
at 6~M, and the corresponding film displays an extremely flat surface. 
Layer-by-layer growth is then observed up to a dozen of monolayers, 
with highly flat morphologies occurring at even-layer Pb coverage. At 
larger thickness, bi-layer modulations dominate the HAS reflectivity.
The analysis of the XRD rod scans suggests that the Pb pseudomorphic layer 
is decomposed to contribute to the Pb(111) film. In fact, 
the number of 
Pb(111) layers within the film corresponds to the number of deposited 
Pb monolayers. On the contrary, the pseudomorphic layer produced by 
LT deposition on Si(111) has been found by XRD to be preserved also 
below a thick Pb film  \cite{howes}.

If the temperature is not low enough, or the substrate terrace size is 
not large enough, island growth sets in and the height correlation between 
different islands is lost, thus smearing the layer-by-layer features.
For comparison, we recall that Pb islands are also observed for 
Pb deposition on Si(111) when the substrate temperature exceeds 
$\sim$~150~K \cite{budde,hupalo1,hupalo2}, while layer-by-layer growth 
is achieved from 100~K \cite{schmicker} down to 16~K \cite{jalochowski3}.
The 2D films grown on Ge(001) 
are metastable and their long range order can be increased 
by short annealing 
to $\sim$~190~K after LT Pb deposition. Upon annealing at higher temperature,  
the films are always irreversibly 
decomposed by fragmentation into uncorrelated 3D islands
at any 
Pb film thickness. Metastability is also observed for Ag films 
on GaAs(110) at thickness higher than 6~ML. In this case 
the flat Ag islands formed by LT deposition are 
seen to coalesce into a flat wetting film upon annealing to room 
temperature \cite{yu}. 
Further annealing to 670~K irreversibly decomposes the 
Ag film into 3D mound-like islands. 

The ``electronic growth'' model qualitatively describes the LT 
layer-by-layer growth of Pb on Ge(001), but a detailed calculation of 
the interface energetics would be needed to explain the observed 
larger stability of the even-layer Pb films (odd layer stability is 
predicted for Pb atoms \cite{zhang}). An atomistic model of 
the diffusion and growth mechanism is still missing. The 
mechanism of funneling \cite{jevans}, proposed for the 
Pb/Si(111) system \cite{jalochowski3,schmicker}, 
does not seem to work in the present case (not exclusively, at least). 
In fact, the absence of 
surface diffusion would lead to a high island density, that, even if 
smoothed by the funnelling effect, would yield a large density of 
point defects, as a consequence the HAS reflectivity would be 
strongly reduced \cite{nota4}. On the contrary the HAS reflectivity is extremely 
high (two order of magnitude higher than on the Ge(001) surface) and 
the  Pb(111) domains display a mean size of several hundreds of \AA ngstroms.

Quantum size effects have been observed to affect both the surface 
charge density and the surface relaxation of the growing Pb film.
The strongest effect is the apparent variation of the step 
height as measured by HAS \cite{crottini}. 
In fact, the He turning point at 
the Pb(111) surface is displaced along the surface normal 
according to the in-vacuum 
decay of the electronic charge. 
Adjacent terraces of different 
height level display different decays of their surface charge density since 
they accommodate a different number of QWS \cite{su}. The
apparent step height, as measured by both HAS and STM, is thus 
affected since these investigation techniques  
are only sensitive to the surface charge density.
The displacement of the charge density drives the displacement of 
the ionic cores, thus affecting the 
surface relaxation \cite{feibelman,batra}.
By XRD experiments, the extent of the structural QSE has been measured. 
Step height variations up to 5-6~\%~ have been determined, as opposed to 
 20~\%~ obtained by the HAS measurements. Although smaller 
than the displacement of the He turning point, the displacement of the 
topmost nuclear plane is non negligible, as previously thought 
\cite{jalochowski3,braun,materzanini}. Most importantly, the topmost 
layer effectively follows the direction of the charge density 
displacement, thus opening the way to the manifestation of other QSEs.

A sizeable variation of the surface relaxation may lead to the 
variation of the elastic force constant between 
the first and second topmost layers, thus affecting the acoustic 
vibrations. The dispersion of low energy 
surface phonons has been measured by inelastic HAS at different Pb 
film thickness. 
Appreciable differences among the phonon frequencies have been 
detected throughout the $\Gamma K$ and $\Gamma M$ direction of the 
first surface Brillouin zone. However, these differences could not be 
unambiguously correlated with the observed cyclic relaxations of the 
Pb topmost layer.

\section{Acknowledgments}

We are indebted with Fernando Tommasini who designed, realized and 
made working both the HAS apparatus and the ALOISA beamline.
This project was partly funded by MURST cofin99 (Prot. 9902112831)




\begin{thebibliography}{00}
\bibitem{nota1} This mixed phase was erroneously identified with a 
$(4 \times 1)$ symmetry phase in Ref.~\cite{crottini}.
\bibitem{nota2} It should be noted that the deposition curve published in 
Fig. 1 of Ref.~\cite{crottini} was not normalized to a constant deposition rate 
nor calibrated. In fact, the consecutive peak labelling of Ref.~\cite{crottini} 
is always referred to the deposition stages, which are easily 
identified with the corresponding ones shown in the present paper.
\bibitem{nota3} The 
Pb bulk lattice parameter is 4.95 \AA, and the distance between 
equivalent rows on the (111) surface is $3.03$ \AA.
\end{thebibliography}
\end{document}